\documentclass[useAMS,usenatbib,usegraphicx]{mn2e}

\def\simlt{\lower.5ex\hbox{$\; \buildrel < \over \sim \;$}}
\def\simgt{\lower.5ex\hbox{$\; \buildrel > \over \sim \;$}}

\def\beq{\begin{equation}}
\def\eeq{\end{equation}}

\def\Omm{{\Omega_{\rm m}}}
\def\Ommz{{\Omega_{\rm m}^{\,z}}}
\def\Oml{{\Omega_{\Lambda}}}
\def\tE{\theta_{\rm E}}

\def\zL{z_{\rm L}}
\def\zS{z_{\rm S}}
\def\c2{c_{200}}
\def\cv{c_{\rm vir}}
\def\r2{r_{200}}
\def\rv{r_{\rm vir}}
\def\rs{r_{\rm s}}
\def\Mv{M_{\rm vir}}
\def\M2{M_{200}}
\def\arcmin{^\prime}
\def\arcsec{^{\prime\prime}}

\title{Large Einstein Radii: A Problem for $\Lambda$CDM}
\author[Tom J.\ Broadhurst and Rennan Barkana]{Tom J.\
Broadhurst$^{1}$ and Rennan Barkana$^{2,3}$\thanks{E-mail:
tjb@wise.tau.ac.il; barkana@wise.tau.ac.il}\\ $^{1}$Raymond and
Beverly Sackler School of Physics and Astronomy, Tel Aviv University,
Tel Aviv 69978, Israel\\ $^{2}$ Institute for Cosmic Ray Research,
University of Tokyo, Kashiwa 277-8582, Japan \\ $^{3}$ Guggenheim
Fellow; on sabbatical leave from the School of Physics and Astronomy,
Tel Aviv University, Israel}

\begin{document}

\pagerange{\pageref{firstpage}--\pageref{lastpage}} \pubyear{2008}
\maketitle
\label{firstpage}
\begin{abstract}

The Einstein radius of a cluster provides a relatively
model-independent measure of the mass density of a cluster within a
projected radius of $\sim 150$ kpc, large enough to be relatively
unaffected by gas physics. We show that the observed Einstein radii of
four well-studied massive clusters, for which reliable virial masses
are measured, lie well beyond the predicted distribution of Einstein
radii in the standard $\Lambda$CDM model.  Based on large samples of
numerically simulated cluster-sized objects with virial masses $\sim
10^{15}M_{\odot}$, the predicted Einstein radii are only
$15-25\arcsec$, a factor of two below the observed Einstein radii of
these four clusters. This is because the predicted mass profile is too
shallow to exceed the critical surface density for lensing at a
sizable projected radius. After carefully accounting for measurement
errors as well as the biases inherent in the selection of clusters and
the projection of mass measured by lensing, we find that the
theoretical predictions are excluded at a 4-$\sigma$
significance. Since most of the free parameters of the $\Lambda$CDM
model now rest on firm empirical ground, this discrepancy may point to
an additional mechanism that promotes the collapse of clusters at an
earlier time thereby enhancing their central mass density.

\end{abstract}


\section{Introduction}\label{intro}

The standard picture of the basic cosmological framework has recently
come to rest firmly on detailed empirical evidence regarding the
cosmological parameters, the proportions of baryonic and non-baryonic
dark matter, together with the overall shape and normalization of the
power spectrum \citep[e.g.,][]{SNV06,Spergel07,BAO07}.  This framework
has become the standard $\Lambda$CDM cosmological model, with the
added simple assumptions that the dark matter reacts only to gravity,
is initially sub-relativistic, and possesses initial density
perturbations which are Gaussian distributed in amplitude. This is a
very well defined and relatively simple model, with clear predictions
which are amenable to examination with observations. The cooling
history of baryons complicates the interpretation of dark matter on
galaxy scales, especially for dwarf galaxies that traditionally have
been a major focus of studies of halo structure. Clusters have the
advantage that the virial temperature of the associated gas is too hot
for efficient cooling, so the majority of the baryons must trace the
overall gravitational potential and hence we may safely compare
lensing-based cluster mass measurements to theoretical predictions
that neglect gas physics and feedback.

Lensing-based determinations of the mass profiles of galaxy clusters
rely on detailed modeling of the strong lensing region to define the
inner mass profile, and also a careful analysis of the outer weak
lensing regime. The latter involves substantial corrections for
instrumental and atmospheric effects \citep{KSB}, and a clear
definition of the background, free of contamination by the lensing
cluster \citep{Br05b,Medezinski}. In the center we may make use of the
Einstein radius of a cluster which is often readily visible from the
presence of giant arcs and provides a relatively model-independent
determination of the central mass density. In the case of axial
symmetry, the projected mass inside the Einstein radius $\tE$ depends
only on fundamental and cosmological constants: $M(<\tE)=\tE^2
(c^2/4G) D_{\rm OL} D_{\rm OS} / D_{\rm LS}$, where this combination
of angular diameter distances (observer-lens, observer-source, and
lens-source) leads to a relatively weak dependence on the lens and
source redshifts. More generally, an effective Einstein radius can be
defined by axially averaging the projected surface density, which
itself is well determined when there are a large number of
constraints. Virtually all known massive clusters at intermediate
redshifts, $0.15<z<0.8$, show multiple images including obvious arcs
in sufficiently deep high-resolution data. The derived Einstein radius
of these massive clusters typically falls in the range $10\arcsec<
\tE< 20\arcsec$ \citep{Gioia,Smith}, with the largest known case of
$\sim 50\arcsec$ for A1689.

Increasingly large simulations have helped to specify the evolution of
the halo mass function and the form of the mass profile predicted in
the context of the $\Lambda$CDM model \citep{NFW,Bull}. These
simulations are now becoming sufficiently large and detailed to define
the predicted spread of halo structure over a wide range of halo mass,
and to quantitatively assess the inherent bias in observing clusters
in projection and selecting them by lensing cross-section
\citep{Hennawi,Neto}. For the most massive collapsed objects in these
simulations (virial mass $\Mv \sim 10^{15} M_{\odot}$), a mean
observed concentration of $\c2\sim 6$ is predicted for lenses, where
$\c2$ is defined precisely in the next section. Such profiles are
relatively shallow and seem at odds with recent careful lensing
studies of massive clusters; although the \citet{NFW} (hereafter NFW)
profile provides acceptable fits to the observations, relatively high
concentrations of $\cv\sim 10-15$ are derived for several well-studied
massive clusters
\citep{Kneib,Gavazzi,Br05b,kling,Limousin,Bradac,halkola08,umetsu}. These
values are larger than expected based on simulations of the standard
$\Lambda$CDM model. Given the relatively shallow mass profile
predicted for cluster-mass CDM halos, the question arises whether the
projected critical surface density for lensing can be exceeded within
a substantial radius for this model.

In this paper we compare observations of well-constrained massive
clusters with the predictions of $\Lambda$CDM simulations. The idea is
to compare directly the projected 2-D mass distributions in the
observations and the simulations. In the observations, the projected
surface density is obtained from the lensing analysis, which we
emphasize does not assume any symmetry of the lensing mass. In the
simulations, the 2-D density is directly measurable from the
simulation outputs. We summarize each 2-D density distribution with
two defined quantities, the virial mass and the effective Einstein
radius. In the simulations, the 2-D density distribution of each halo
was already axially averaged by fitting to a projected NFW profile, so
we use this profile to obtain the effective $\tE$ (which,
consistently, is also defined through an axial average). The virial
mass is measured directly in the simulations (with its usual
spherically-averaged definition). In the observations, we obtain $\Mv$
directly in A1689 (though in projection), and measure it using NFW
fits to the final distribution in the other clusters.

This paper is structured as follows. In section~2 we first summarize
the theoretical predictions, including a brief review of the NFW
profile and its lensing properties, and of the halo concentrations
measured by \citet{Neto} and \citet{Hennawi} in large numerical
simulations. Note that the conflict between high observed
concentrations and lower ones determined for the numerical halos was
noted in both of these papers (see also \citet{navarro}). We then
present the observational data for the four clusters, followed by a
model-independent method for measuring the mass, which we apply to
A1689. In section~3 we confront the theoretical predictions with the
data, finding a clear discrepancy. We discuss the possible
implications in section~4.

\section{Theoretical and Observational Inputs}

\subsection{Theoretical Predictions}

Our calculations are made in a cold dark matter plus cosmological
constant (i.e., $\Lambda$CDM) universe matching observations
\citep{Spergel07}, with a power spectrum normalization
$\sigma_8=0.826$, Hubble constant $H_0=100h \mbox{ km s}^{-1}\mbox{
Mpc}^{-1}$ with $h=0.687$, spectral index $n = 0.957$, and present
density parameters $\Omega_m=0.299$, $\Omega_\Lambda=0.701$, and
$\Omega_b=0.0478$ for matter, cosmological constant, and baryons,
respectively. Unless otherwise indicated, we use physical units that
already include the proper factors of $h$ or $h^{-1}$, always with
$h=0.687$.

Consider a halo that virialized at redshift $z$ in a flat universe
with $\Omm+\Oml=1$. At $z$, $\Omm$ has a value \beq \Ommz= \frac{\Omm
(1+z)^3}{\Omm (1+z)^3+\Oml}\ , \label{Ommz} \eeq and the critical
density is \beq \rho_{\rm c}^z = \frac{3 H_0^2} {8 \pi G} \frac{\Omm
(1+z)^3}{\Ommz} \ . \eeq The mean enclosed virial density in units of
$\rho_{\rm c}^z$ is denoted $\Delta_c$ and used to define the virial
mass and radius in observations and in simulations. Sometimes a fixed
value is used, such as $\Delta_c = 200$, although the theoretical
value is $\Delta_c=18\pi^2 \simeq 178$ in the Einstein-de Sitter
model, modified in a flat $\Lambda$CDM universe to the fitting formula
\citep{bn98} \beq \Delta_c=18\pi^2+82 d-39 d^2\ , \label{eq:theory} \eeq
where $d\equiv \Ommz-1$. A halo of mass $M$ collapsing at redshift $z$
thus has a (physical) virial radius \beq \rv=1.69
\left(\frac{M}{10^{15}\ M_{\sun} }\right)^{1/3} \left[\frac{\Omm h^2}
{\Ommz}\ \frac{\Delta_c} {18\pi^2}\right]^{-1/3} \frac{1}{1+z}\ {\rm
Mpc}\ . \label{rvir}\eeq

Numerical simulations of hierarchical halo formation indicate a
roughly universal spherically-averaged density profile for virialized
halos \citep{NFW}, though with considerable scatter among different
halos \citep[e.g.,][]{Bull}. The NFW profile has the form \beq
\rho(r)=\rho_{\rm c}^z\, \frac{\delta_c} {x (1+x)^2}\ , \eeq where
$x=r/\rs$ in terms of the NFW scale radius $r_s = \rv/\cv$, and the
characteristic density $\delta_c$ is related to the concentration
parameter $\cv$ by \beq \delta_c= \frac{\Delta_c}{3} \frac{\cv^3}
{\ln(1+\cv)-\cv/(1+\cv)} \ . \eeq For a halo of mass $M$ at a given
redshift $z$, the profile is fixed once we know $\Delta_c$ and
$\cv$. In this paper we denote the concentration parameter, virial
radius and mass by $\cv$, $\rv$ and $\Mv$ when using the theoretical
value in equation~(\ref{eq:theory}), and by $\c2$, $\r2$, and $\M2$,
respectively, when using $\Delta_c = 200$.

The lensing properties of a halo are determined by $\kappa$, the
projected surface density $\Sigma$ measured in units of the critical
surface density $\Sigma_{\rm cr}=[c^2/(4\pi G)]D_{\rm OS}/(D_{\rm OL}
D_{\rm LS})$. For an axisymmetric lens, the Einstein radius (i.e.,
tangential critical curve) occurs at a projected radius $R$ where the
mean enclosed surface density satisfies $\bar{\kappa}(R)=1$. For an
NFW halo, letting $X=R/\rs$ we have $\bar{\kappa}(X)=(4/\Sigma_{\rm
cr}) \rho_{\rm c}^z \delta_c \rs\, g(X)/X^2$ where \citep{Bart96} \beq
g(x)=\ln{x\over2} + \cases{ 1\ , & $x=1$ \cr {2\over\sqrt{x^2-1}}{\rm
tan}^{-1}\sqrt{x-1\over x+1}\ , & $x>1$ \cr {2\over\sqrt{1-x^2}}{\rm
tanh}^{-1}\sqrt{1-x\over 1+x}\ , & $x<1$ \cr }\;. \eeq

We derive the theoretical predictions for cluster lensing in
$\Lambda$CDM by combining the two largest studies of halo structure in
cosmological numerical simulations. \citet{Neto} studied halo
structure within the Millennium simulation, in which over 2000 halos
formed with $\M2>10^{14} M_\odot$ at $z=0$. Each halo was resolved
with $>80,000$ particles, allowing a detailed look at its
three-dimensional density structure using NFW profile fitting. This is
therefore the best available statistical analysis of the cluster halo
population in $\Lambda$CDM simulations. However, since \citet{Neto}
did not study projected halo profiles or the bias in selecting lenses,
we must adjust their results in order to apply them to
lensing. \citet{Hennawi} studied 900 simulated cluster halos at
$z=0.41$, each resolved into at least $30,000$ particles. They studied
projections through on average 15 random directions per cluster (more
-- up to 125 -- for the most massive ones), and fit both 3-D and
projected 2-D NFW profiles. Moreover, they separately studied the
distribution of NFW profile parameters both for the general halo
population and for the lensing population (i.e., where halos are
weighted by their strong lensing cross section). They showed that the
inherent triaxiality of CDM halos along with the presence of
substructure enhance the projected mass in some orientations, leading
to a bias in the 2-D structure of lenses compared with the 3-D
structure of the general population of cluster halos (see also
\citet{Oguri}).

\begin{figure}
\includegraphics[width=84mm]{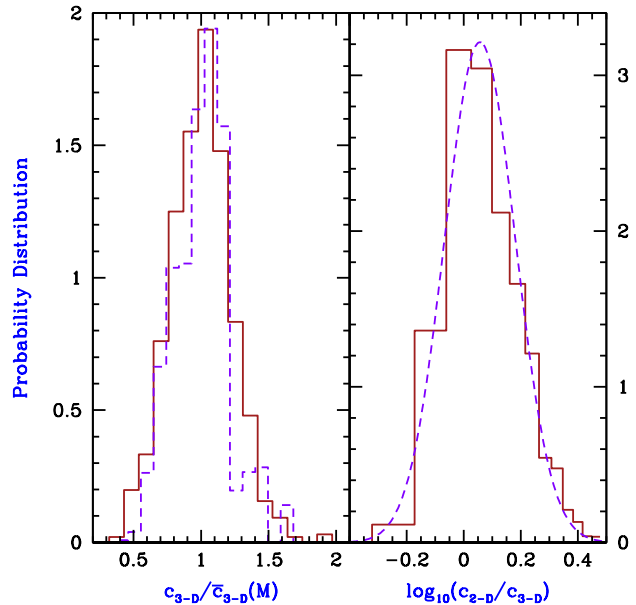}
\caption{Probability distributions of various concentration
parameters, based on \citet{Hennawi}. {\it Left panel:}\/ When the
concentration parameters are measured relative to the median at each
halo mass (Figure~8 of \citet{Hennawi}), the distribution for the
general population (solid histogram) matches that for the lens
population when the latter is divided by a factor of 1.17 (dashed
histogram). {\it Right panel:}\/ For lensing halos, the distribution
of $\log_{10}$ of the ratio of the 2-D to the 3-D concentration (solid
histogram) is well fitted by a Gaussian with the same mean and
variance as the histogram (dashed curve). Note that the non-uniform
binning is a result of our conversion of the linear-axis histogram
from Figure~12 of \citet{Hennawi} to one with a logarithmic $x$-axis.}
\label{fig:Hennawi}
\end{figure}

Figure~\ref{fig:Hennawi} shows various biases in halo concentration
parameters $\cv$ as measured by \citet{Hennawi}. The Figure shows
(left panel) that the distribution of 3-D concentrations of the lens
population is the same as that of the general halo population except
for a shift upwards by a factor of 1.17. For a given real 3-D profile,
the 2-D profile measured in projection depends on the orientation, and
is thus given by a probability distribution. The Figure also shows
(right panel) that the ratio $c_{\rm 2-D}/c_{\rm 3-D}$ follows a
lognormal distribution, i.e., that $\log_{10}$ of this ratio is well
fitted by a Gaussian with a mean value of 0.057 and $\sigma=0.124$,
which correspond to factors of 1.14 and 1.33, respectively.

At $z=0.41$, $\r2$ and $\c2$ are typically $\sim 15\%$ smaller than
$\rv$ and $\cv$, respectively, and $\M2 \approx 0.9 \Mv$. We assume
that the {\it relative}\/ distributions shown in
Figure~\ref{fig:Hennawi} are approximately the same for $\c2$ as for
$\cv$, and that they are independent of halo mass within the narrow
range considered, as found by \citet{Hennawi}. Thus, we can apply
these findings to convert the 3-D $\c2$ distributions measured by
\citet{Neto} in order to obtain the resulting 2-D projected
distributions of $\c2$ values for the population of cluster
lenses. Specifically, we multiply the $\c2$ values by a factor of 1.17
(the lensing bias) and then convolve with the distribution of $c_{\rm
2-D}/c_{\rm 3-D}$ for lenses (the projection bias).  Note that we use
the values measured by \citet{Neto} at $z=0$ when comparing to the
observed clusters at redshifts $z=0.18-0.40$. This is a conservative
assumption, since the typical concentration parameter at a given halo
mass declines with redshift. Studies based on large numerical
simulations \citep{Jing,Gao} found for massive halos a relatively weak
decline of $\sim 20\%$ out to $z=1$, which suggests a $5-10\%$ decline
out to the redshifts we consider below, a decline which we do not
include here.
 
\begin{figure}
\includegraphics[width=84mm]{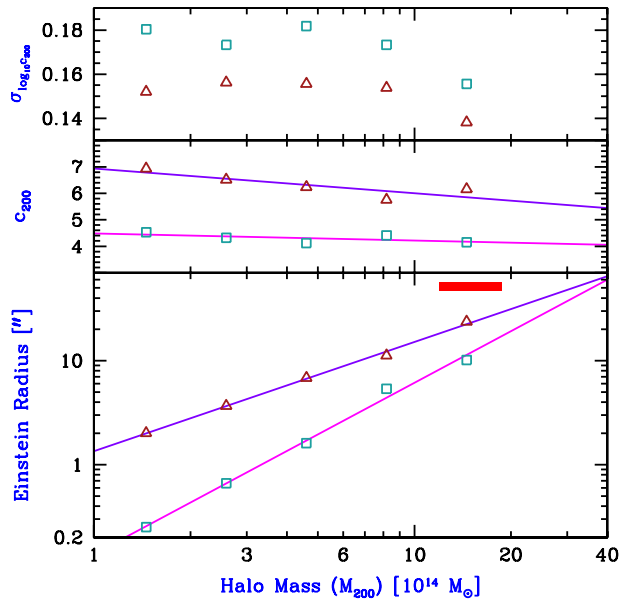}
\caption{Dependence of $\tE$ (bottom panel), the median $\c2$ (middle
panel), and the scatter $\sigma_{\log_{10}\!\c2}$ (top panel) on halo
mass $\M2$ in numerical simulations. We use the 3-D analysis by
\citet{Neto} corrected for lensing and projection bias based on
Figure~\ref{fig:Hennawi}. We consider the relaxed (triangles) or
unrelaxed (squares) halo populations. We assume the median $\c2$ and
the A1689 redshifts when calculating $r_E$. We show several linear
least-squares fits to help discern trends (solid curves). Also shown
for comparison (narrow box, bottom panel) is the location
corresponding to the observations of A1689 (where the boxed area
contains the two-sided 1-$\sigma$ ranges of $\M2$ and $\tE$).}
\label{fig:NetoMean}
\end{figure}

In Figure~\ref{fig:NetoMean} we show the predicted Einstein radii of
$\Lambda$CDM cluster halos of mass $\M2 > 10^{14} M_\odot$ based on
\citet{Neto} and \citet{Hennawi}. \citet{Neto} divided their halos
within each mass bin into ``relaxed'' and ``unrelaxed'' groups of
halos, the latter identified as being disturbed dynamically as
indicated by a large amount of substructure, a large offset between
the center of mass and the potential center, or a high kinetic energy
relative to potential. For each group, they found that the statistical
distribution of the concentration parameters was well-fitted by a
lognormal distribution. Thus, after correcting for lensing and
projection bias as explained above, the resulting distributions remain
lognormal. As shown in Figure~\ref{fig:NetoMean}, the median $\c2$ of
each group is only weakly dependent on mass, showing a slight trend of
decreasing concentration with increasing halo mass (a trend which is
more apparent over the broader range considered by \citet{Neto}, down
to $\M2 = 10^{12}M_\odot$). As a result, we find that $\tE \propto
\M2$ (relaxed) and $\tE \propto (\M2)^{1.6}$ (unrelaxed). Note that for
a lognormal distribution, the median and mean are theoretically the
same if $\log_{10} \c2$ is considered rather than $\c2$. As the Figure
shows, the typical scatter in $\log_{10} \c2$ among halos of the same
mass is also fairly independent of mass, except for the highest mass
bin, $\M2 > 10^{15} M_\odot$. While this bin is based on a somewhat
small sample (8 relaxed and 11 unrelaxed halos),
\citet{Neto} suggest that the lower dispersion is expected since the
highest-mass halos are very rare, thus all formed very recently and
should have similar merger histories and thus internal
structures. Note that the final, effective scatter in our calculations
is only $\sim 10\%$ lower for the last bin compared with the lower
mass bins, since we assume that the projection scatter is independent
of halo mass.

\subsection{Observational Data}

For our data set of lensing clusters we choose four well-studied
clusters with strong constraints available both from multiply-imaged
arcs in the strong-lensing regime and from distorted arcs and
magnification measurements in the weak-lensing regime. We determine an
effective Einstein radius in each cluster using the 2-D projected mass
distribution obtained from fitting (without assuming symmetry) to the
large number of sets of multiple images; from the obtained $\kappa$ we
then define $\tE$ in the standard way, as the radius enclosing a mean
surface density equal to the critical surface density for
lensing. Note that this definition effectively axially averages the
mass distribution. We can define $\tE$ relative to any source
redshift, and in each cluster we choose a fiducial redshift that
matches a prominent arc system. Our procedure for determining $\tE$ is
also a good match to that followed by \citet{Hennawi} for their
simulated clusters, where they fit an axially symmetric model to the
projected mass distribution; we obtain the predicted effective
Einstein radii of the simulated clusters based on their model fits, as
detailed in the previous subsection. For the observed clusters, we
also obtain effective virial masses from combined axially-symmetric
fits to the strong and weak lensing data.

Very deep HST/ACS imaging of several massive lensing clusters has been
obtained by the ACS/GTO team \citep{Ford}. This includes the
well-studied cluster A1689 $(z=0.183)$ for which several comprehensive
strong lensing analyses have been published. The analysis of
\citet{Br05a} identified over 100 multiply lensed images of 30
background galaxies by an iterative procedure in which multiple images
are securely identified by delensing and relensing background
galaxies. These multiple images have been used as a basis for other
types of modeling, including fully parametric
\citep{Halkola,Zekser,Limousin} and non-parametric modeling
\citep{Diego}. \citet{Br05b} combined these data on the inner mass
profile with wide field archival Subaru images and obtained a detailed
radial mass profile for the entire cluster, finding an NFW-like
projected mass profile with a surprisingly high value for the
concentration parameter ($\cv=13.7\pm1.2$), and a virial mass of
$\Mv=(1.9\pm0.2)\times 10^{15}M_{\odot}$ (the value in Table~1 is
different; see the next subsection). This yields an effective Einstein
radius of $52\arcsec$ at a fiducial redshift $z=3$, in good agreement
with the radius of $\sim 50\arcsec$ found for prominent multiple arcs
at this redshift. We adopt a conservative error estimate of $10\%$ on
$\tE$ for all the clusters (Table~1). Note that an Einstein radius of
only $31\arcsec$ is implied by the NFW fit of \citet{Limousin} to
independent weak lensing data from CFHT, but this number is
incompatible with their own strong lensing analysis and is possibly
caused by contamination of the lensing signal by unlensed cluster
members; if not thoroughly excluded, this contamination will reduce
the lensing signal preferentially towards the cluster center,
resulting in a flatter mass profile, as pointed out by \citet{Br05b}.

\begin{table}
 \centering
  \caption{Observational Data}
  \begin{tabular}{@{}lrrrrr@{}}
  \hline
  Cluster &
  $\Mv$ [$M_\odot$] &
  $\M2$ [$M_\odot$] &  
  $\tE$ [$\arcsec$] &
  $\zL$ &  $\zS$\\
 \hline
A1689 & $1.6 \times 10^{15}$ & $1.5 \times 10^{15}$ & 52 & 0.183 & 3 \\
Cl0024-17 & $8.7 \times 10^{14}$ &  $8.0 \times 10^{14}$ & 31 & 0.395 & 1.7 \\
A1703 & $1.0 \times 10^{15}$ &  $9.0 \times 10^{14}$ & 32 & 0.258 & 2.8 \\
RXJ1347 & $1.3 \times 10^{15}$ &  $1.2 \times 10^{15}$ & 35 & 0.45 & 1.8 \\
\hline
\end{tabular}
\end{table}

Another well studied cluster that we use here is Cl0024-17
($z=0.395$), with an effective Einstein radius of $31\arcsec$ at
$z=1.7$ defined by several sets of multiple images identified in the
ACS/GTO images. This agrees in particular with the mean radius of the
famous 5-image system of ``$\theta$'' arcs at a spectroscopically
measured redshift of $z=1.685$ \citep{Br00}. This set of multiple
images and the distortion measurements of background galaxies with
photometric redshifts has been used by \citet{Jee} to constrain the
inner mass profile. Their result is in general consistent in form with
the earlier analysis of the inner profile by \citet{Br00}, but with
the addition of a narrow low-contrast ring that, it is claimed, can be
reproduced in simulations where merging of two massive clusters occurs
along the line of sight. A line-of-sight merger is also blamed for the
relatively small central velocity dispersion \citep{Czoske}. However,
in the extensive weak lensing analysis of \citet{Kneib} only one small
subgroup is visible, offset by $3\arcmin$ in projection from the
center of mass and accounting for only $\sim 15\%$ of the total mass
of the cluster. \citet{Kneib} find that the main cluster is well
fitted by an NFW profile with a virial mass of $\sim
6\times10^{14}M_{\odot}/h$ and with a high concentration, $\cv \sim
20$. In a more recent analysis of deep multicolour B,R,Z Subaru
images, Medezinski et~al.\ (2007, in preparation) find good agreement
with \citet{Kneib} (Table~1).

In addition, we use the very deep ACS/GTO images of the massive
cluster A1703 for which many sets of multiple images are visible, so
that the tangential critical line is easily identified with a mean
Einstein radius of $32\arcsec$ at $z=2.8$, in good agreement with the
radius of the main giant arc at a similar redshift (Table~1). In the
weak lensing analysis of Medezinski et~al.\ (2007, in preparation) a
very good fit to an NFW profile is found with a virial mass $\Mv=
7\times10^{14} M_{\odot}/h$. This cluster appears relaxed and
centrally concentrated, with little obvious substructure. To date,
deep X-ray imaging is unfortunately missing.

Finally, a weak lensing analysis of RXJ1347 (Medezinski et~al.\ 2007,
in preparation) shows this cluster to have a very circular shear
pattern, with an estimated virial mass $\Mv= 9\times10^{14}
M_{\odot}/h$ based on an NFW fit to the radial distortion
profile. This cluster has the highest observed X-ray temperature of 13
keV and a symmetric X-ray emission map that indicates that it is
relaxed \citep{vik}. A very symmetric distribution of arcs is visible
around the cluster center, implying a well-determined Einstein radius
at $z=1.8$ of $35\arcsec$ from the full model, a value which is also
in agreement with a system of 5 multiply-lensed images at this
redshift \citep{halkola08}.

These four clusters are particularly useful for our purpose, by virtue
of their well-defined Einstein radii and precise measurements of the
virial masses, which allows a comparison with the theoretical
predictions as a function of halo mass. We convert $\Mv$ to $\M2$ for
each cluster using the measured value of $\cv$, and adopt error bars
of $\pm 15\%$ on $\M2$ and $\pm 10\%$ on $\tE$ for all four clusters
(but see the next subsection for an alternative measurement of the
virial mass of A1689). It is also interesting to note the many
examples of strong lensing by other galaxy clusters for which the
total mass is not so well constrained. Samples of clusters defined by
some reasonable criteria \citep{Smith,Sand,Comerford} show that
invariably the observed Einstein radius (when detected) for
intermediate redshift clusters does not fall short of $10\arcsec$,
with a mean of $\sim 15\arcsec$. This may be compared with the
predicted typical Einstein radius of only $\sim 5\arcsec$ from the
simulations of \citet{Neto} and \citet{Hennawi} for a cluster of
$\M2=$~several~$\times 10^{14} M_\odot$ (Figure~\ref{fig:NetoMean}).

\subsection{Model-Independent Mass}

Of the two observational quantities we use to characterize each
cluster, $\tE$ is more directly estimated, from the positions of
multiple images. The mass $\M2$ requires a measured mass profile out
to large angles, which can be used to estimate the angular position
corresponding to $\r2$, i.e., to an enclosed relative density of 200
times the critical density. Deep images provide a large density of
weakly-lensed background sources, but weak lensing distortions measure
only the reduced shear and suffer from the well-known mass-sheet
degeneracy. This means that the mass profile can be measured without
degeneracies only by fitting a particular parametrized density
profile model to the data. However, combining lensing distortions with
observations of the variation in the number density of background
sources due to weak magnification breaks the degeneracy and yields a
direct measurement of the projected surface density in each radial bin
\citep{Br05b}. Given such independent measurements out to large
radius, we can derive the corresponding value of $\M2$ directly from
the data, without the intermediary of an assumed model profile, the
use of which inevitably introduces a non-trivial systematic
error. Such accurate measurements are available for A1689, which we
use to illustrate the method, and such data should be obtainable for
other clusters as well. We note that this effective virial mass is
defined from deprojecting the projected mass assuming spherical
symmetry, which is the closest lensing observations can come to the
standard theoretical definition of the virial mass based on a 3-D
spherical average.

Lensing by a halo can be analyzed by calculating
$\kappa=\Sigma/\Sigma_{\rm cr}$. The projected surface density is
related to the three-dimensional density $\rho$ by an Abel integral
transform. This implies a relation between the integrated
three-dimensional mass $M(r)$ out to radius $r$ and $\kappa(R)$ as a
function of the projected radius $R$: \beq M(r) =
\Sigma_{\rm cr} \left[ 2 \pi \int_0^r R\, \kappa(R)\, dR -4
\int_r^\infty R\, \kappa(R) f\left( \frac{R}{r} \right)\, dR \right]\
, \label{eq:M} \eeq where \beq f(x)= \frac{1} {\sqrt{x^2-1}} - {\rm
tan}^{-1} \frac{1} {\sqrt{x^2-1}}\ . \eeq The first term in
equation~(\ref{eq:M}) is the total projected mass within a ring of
projected radius $r$, and the second term removes the contribution
from mass elements lying at a 3-D radius greater than $r$.

To obtain the 3-D mass profile $M(r)$, we apply equation~(\ref{eq:M})
to the 26 values of $\kappa(R)$ measured by \citet{Br05b} over the
range $R=0.015-2.3$ Mpc in A1689. Specifically, we linearly
interpolate $\kappa(R)$ between each pair of measured points, and
extrapolate outside the range. We extrapolate inward assuming
$\kappa(R)={\rm const}$ from the innermost point and outward assuming
$\kappa(R) \propto R^{-2}$ from the outermost point, where these power
laws are motivated by the NFW profile. However, even varying these
powers by $\pm 1$ would change the virial mass by only $1\%$, which is
negligible compared with the effect of the measurement errors. Once we
have obtained $M(r)$ at all $r$, we interpolate to find the
appropriate $r$ that yields a desired mean enclosed density, and thus
determine the virial radius and mass. The error analysis is
complicated by the fact that the virial radius is not fixed but rather
is itself determined by the data. Thus, to ensure self-consistent
errors we use a Monte Carlo approach, generating 1000 random profiles
of $\kappa(R)$ according to the measurement errors (assumed to be
Gaussian distributed except that $\kappa$ is constrained to be
non-negative). For each $\kappa(R)$ profile we find the resulting
virial quantities, and then find the $16\%$, $50\%$ (median), and
$84\%$ percentiles. The result of this direct, model-independent
analysis of A1689 is: $$ \rv = 2.76\pm0.2\ {\rm Mpc}, \ \ \Mv = (1.6
\pm 0.4) \times 10^{15} M_\odot, $$ $$ \r2 = 2.25\pm0.2\ {\rm Mpc}, \
\ \M2 = (1.5 \pm 0.4) \times 10^{15} M_\odot. $$

The median value obtained for $\M2$ is lower by $\sim 15\%$ than the
best-fit value from the NFW profile. This is consistent with our
model-independent analysis of A1689 in \citet{Doron}, where we found
that the NFW fit clearly overshoots the observed density profile at
large radii. Also, the model-independent error of $\sim 25\%$ is
larger since assuming an NFW profile puts a constraint on the density
fluctuations and yields a reduced error. The $25\%$ error results from
allowing completely independent variations in the $\kappa$
measurements at different radii, so the error would be reduced with
even a weak assumption of smoothness in the density profile. Here we
adopt the model-independent mass along with its conservatively large
$25\%$ error (Table~1). The model-independent mass is the most
reliable observationally-determined mass and is more consistent with a
comparison to numerical simulations, where the real virial masses are
known without the need to resort to profile fitting.

The most important assumption in our analysis is spherical
symmetry. Since the measured $\kappa$ profile goes out to $\sim \r2$,
what we directly measure includes the full contribution of $\M2$ plus
additional projected mass coming from larger radii. If the halo has
non-spherical structure such as triaxiality, then lensing bias means
that the contribution of mass elements outside the virial radius to
the projected mass will tend to be unusually high in our
direction. However, since we get that contribution by extrapolating
from the measured $\kappa$ points (which are also enhanced by lensing
bias), the error in our spherical assumption may be small. The real
conclusion is that since the 3-D virial mass is not directly
observable, lensing analyses of halos in numerical simulations should
measure the effective, projected `virial' mass defined by applying
equation~(\ref{eq:M}) to the projected profile. This would allow a
truly direct comparison with the observations.

\section{Confronting $\Lambda$CDM with Observations}

\begin{figure}
\includegraphics[width=84mm]{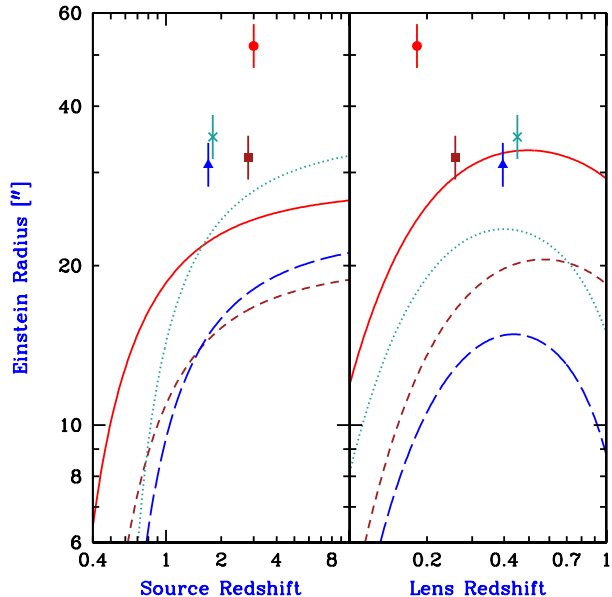}
\caption{Dependence of the Einstein radius $\tE$ on the redshifts
$\zS$ (left panel) and $\zL$ (right panel). We consider A1689 (solid
curves, circles), A1703 (short-dashed curves, squares), Cl0024
(long-dashed curves, triangles), and RXJ1347 (dotted curves,
$\times$'s). In each case, the points correspond to the observed
cluster (with a vertical bar indicating the measurement error), while
the curves show the predicted $\tE$ based on the median $\c2$ of
relaxed simulated halos as measured by \citet{Neto} in the nearest
mass bin, after correction for lensing and projection bias based on
Figure~\ref{fig:Hennawi}.}
\label{fig:rEofz}
\end{figure}

For each cluster with known redshifts $\zL$ and $\zS$ and a reliably
measured value of $\M2$, we can use the results of section~2.1 to
calculate the predicted value of $\tE$ in the $\Lambda$CDM
model. Figure~\ref{fig:rEofz} compares the median expected value with
the observed value for each cluster, showing that in each case the
theoretical expectation falls short of the observed value by about a
factor of two. Specifically, the predicted Einstein radii are
24$\arcsec$, 15$\arcsec$, 16$\arcsec$, and 23$\arcsec$, for A1689,
Cl0024, A1703, and RXJ1347, respectively. In this Figure we also
separately illustrate the dependence of the predicted $\tE$ on the
source and lens redshifts. In general, the predicted $\tE$ increases
with $\zS$, with the dependence weakening as $\zS$ moves further away
from $\zL$. For a fixed $\zS$, $\tE$ is maximized at a particular
value of $\zL$, which for the $\zS$ values of our four clusters falls
at $\zL$ between 0.3 and 0.6. The overall dependence of $\tE$ is weak,
illustrating that the discrepancy between the theory and the
observations would not be significantly affected by small changes in
the redshifts.

\begin{figure}
\includegraphics[width=84mm]{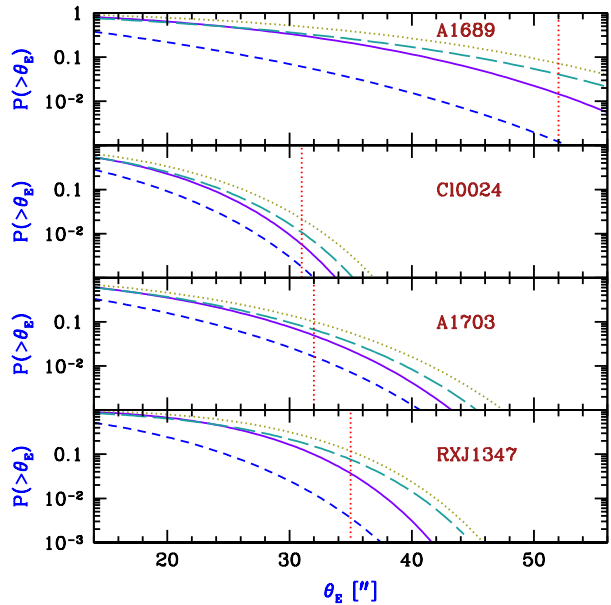}
\caption{Cumulative probability distribution $P(>\tE)$, assuming the
lognormal $\c2$ distribution measured by \citet{Neto} for halos from a
numerical simulation. We consider A1689, Cl0024, A1703, and RXJ1347 as
indicated, assuming in each case the best-fit mass from observations,
and the $\c2$ distribution as measured from the simulation for the
nearest mass bin of relaxed halos (solid curves). We also consider
several possible sources of statistical or systematic error, and we
illustrate the result of assuming a cluster mass higher by the
1-$\sigma$ measurement error (dotted curves), or a scatter
$\sigma_{\log_{10}\!\c2}$ higher by the 1-$\sigma$ sampling noise (see
text; long-dashed curves). In all cases shown, $P(>\tE)$ for unrelaxed
halos would lie below the corresponding curve for relaxed halos,
throughout the plotted region. Thus, we only illustrate the main case
with the $\c2$ distribution as measured for unrelaxed halos
(short-dashed curves). Also shown for comparison for each cluster
(dotted vertical line) is the observed $\tE$.}
\label{fig:NetoDist}
\end{figure}

The results of section~2.1 allow us to make a much more detailed study
of the large Einstein radius problem. We show in
Figure~\ref{fig:NetoDist} the full predicted probability distribution
of $\tE$ for each cluster. The main predictions (solid curves) yield
probabilities of $1.5\%$, $0.56\%$, $5.0\%$, and $3.7\%$ for finding a
cluster with as large a value of $\tE$ (given the redshifts) as
observed for A1689, Cl0024, A1703, and RXJ1347, respectively. In this
Figure and the previous one, we have used the relaxed halo population
from \citet{Neto}. As the Figure shows, including the unrelaxed halos
would only strengthen the inconsistency with the observations. The
results from \citet{Neto} are uncertain due to the relatively small
halo samples used. Specifically, since $\log_{10}\c2$ is Gaussian
distributed, the fractional sampling error with $N \gg 1$ halos is
$1/\sqrt{N}$ in the measurement of the mean of the distribution and
$1/\sqrt{2 N}$ in the standard deviation. Figure~\ref{fig:NetoDist}
shows the result of increasing the standard deviation by its
1-$\sigma$ sampling error; increasing the mean by its 1-$\sigma$ error
would have a smaller effect. While larger simulations will reduce the
sampling noise, the Figure shows that this uncertainty is already
smaller than the effect of the measurement errors in the cluster
masses.

Including the rather conservative observational uncertainties that we
have assumed in $\M2$ and $\tE$, and averaging over Gaussian error
distributions in these two observables, we obtain probabilities of
$8.5\%$, $3.9\%$, $7.9\%$, and $13\%$, for agreement between the
$\Lambda$CDM simulations and A1689, Cl0024, A1703, and RXJ1347,
respectively. If we considered just one of the clusters, the large
Einstein radius problem would only constitute around a 2-$\sigma$
discrepancy. However, we have four independent objects selected from
the population of cluster lenses, and all four are discrepant (in the
same direction). The total probability of the theoretical prediction
yielding four clusters with such large values of $\tE$ is $3 \times
10^{-5}$, which corresponds to a 4-$\sigma$ discrepancy. We emphasize
that we have included in this calculation the lensing and projection
biases, as well as the measurement errors in the cluster masses and
Einstein radii.

\section{Discussion}

We have presented perhaps the clearest, most robust current conflict
between observations and the standard $\Lambda$CDM model. This model
is highly successful in fitting large scale structure measurements,
which in turn strongly constrain the free parameters of the model and
thus produce precise predictions for comparison with data on smaller
scales. Structure on these scales is non-linear and potentially
affected by gas physics, but clusters provide perhaps the best
opportunity for a robust comparison between the models and the
theory. Clusters are so large and massive that their evolution is
dominated by gravity, especially since their high virial temperature
prevents most of the intracluster gas from cooling. The evolution of
clusters including gravitational collapse and virialization can now be
accurately numerically simulated, with sufficient resolution for
studying cluster structure and for simulating lensing in projection,
and in sufficient volumes to produce large samples in a cosmological
context. At the same time, observations of clusters combining weak and
strong lensing now produce accurate virial mass determinations. Given
the virial mass, the cleanest measure of the halo structure is the
effective Einstein radius, which is easily obtained observationally
from a model constrained by large numbers of arcs and directly
measures the central mass density.

We derived the theoretical predictions for cluster lensing in
$\Lambda$CDM using the distribution of 3-D halo profiles measured by
\citet{Neto} in the Millennium simulation, after correcting it for
lensing and projection biases based on \citet{Hennawi}. These analyses
of numerical samples expressed halo structure in terms of the NFW
concentration parameter. We found two key results
(Figure~\ref{fig:Hennawi}) based on the halo analysis by
\citet{Hennawi}: first, that the distribution of 3-D concentrations of
the lens population is the same as that of the general halo population
except for a shift upwards by a factor of 1.17; and second, that the
concentrations measured in projection are related to the 3-D
concentrations, such that the ratio follows a lognormal distribution
which corresponds to a factor of 1.14 shift plus a factor of 1.33
spread.  The concentration parameter is higher for relaxed halos than
for unrelaxed, and it declines slowly with halo mass, resulting in a
predicted Einstein radius that increases roughly linearly with mass
for relaxed halos (Figure~\ref{fig:NetoMean}).

\begin{figure}
\includegraphics[width=84mm]{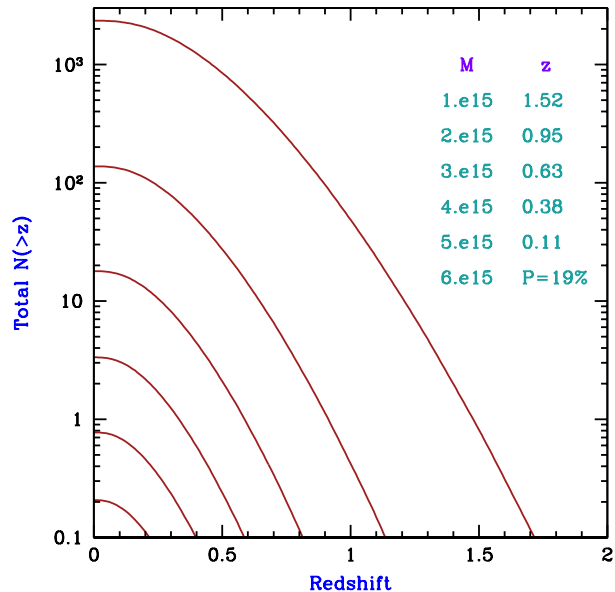}
\caption{Total number of observable clusters in the universe above
redshift $z$, obtained by integrating the halo mass function of
\citet{Sheth} over our past light cone. We consider all cluster halos above
virial mass $M=1$, 2, 3, 4, 5, or $6 \times 10^{15} M_{\odot}$ (top to
bottom). Also listed for each $M$ (top-right corner) is the redshift
above which there is a $50\%$ chance of observing at least one halo of
mass greater than $M$. For $M=6 \times 10^{15} M_{\odot}$ we instead
list the probability of observing at least one halo at any $z>0$.}
\label{fig:Nofz}
\end{figure}

We compared the theoretical predictions with the observed $\tE$ for
four clusters, A1689, Cl0024, A1703, and RXJ1347. For the latter three
we used the virial mass as given by NFW fits to the lensing
observations, but for A1689 we obtained a model-independent mass
directly from the lensing data, only assuming spherical symmetry
(section~2.3). For each object, the predicted $\tE$ values came up
short by a factor of two compared with the observations
(Figure~\ref{fig:rEofz}). After including the measurement errors, the
full probability distribution functions of the predicted Einstein
radii excluded the theoretical model at 2-$\sigma$ for each
object. The total probability of the standard $\Lambda$CDM model
yielding four clusters with such large $\tE$ is $3 \times 10^{-5}$, a
4-$\sigma$ discrepancy.

Lensing work is now being extended to larger samples of clusters, so
that in the near future we may examine more fully the relation between
the Einstein radius and virial mass and its scatter, over a wider
range of cluster masses. The theoretically predicted triaxiality of
CDM halos implies a particular scatter in the projected concentration
parameter (and thus in the Einstein radius) for a given halo
mass. This scatter, which we included in our calculations, is
apparently insufficient to explain the observations. If the scatter is
observationally determined to be relatively small, then this would
further highlight the problem we have discussed and leave the high
concentrations unexplained. Determining the scatter observationally
will also statistically probe the degree of triaxiality of CDM halos.
To ensure the most direct, unbiased comparison, the simulated
distributions should be calculated not at a fixed 3-D virial mass, but
at a fixed projected, effective virial mass, defined as in
section~2.3. We expect this projected virial mass to also be
observationally measured in more clusters.

Numerical simulations show a clear correlation between the
concentration of a halo and its formation time, i.e., the time at
which a significant portion of the halo mass first assembled
\citep[e.g.,][]{Neto}. This agrees with the intuitive notion that a
dense halo core must have assembled at high redshift, when the cosmic
density was high. Thus, the fact that observed cluster halos are
apparently more centrally concentrated than is predicted in
$\Lambda$CDM suggests an additional mechanism that promotes the
collapse of cluster cores at an earlier time than expected. Baryons
are unlikely to help. Central cD galaxies contribute only a small
fraction of the mass within the Einstein radius, which for our four
clusters is $\sim 150$ kpc enclosing a projected mass of $\sim 2
\times 10^{14} M_\odot$, or a 3-D mass of $\sim 1 \times 10^{14}
M_\odot$. We can estimate the effect of baryons on the total mass
profile using the simple model of adiabatic compression
\citep{blumenthal}. Within this model, conservation of angular
momentum implies that the quantity $r M(r)$ (assuming spherical
symmetry) is fixed. Assuming that we start out with a halo with the
mean expected theoretical concentration (Figure~\ref{fig:NetoMean}),
the observed 3-D mass within the Einstein radius can be obtained
through adiabatic compression if the enclosed baryonic mass within
this radius is $\sim 3 \times 10^{13} M_\odot$ in the four clusters we
considered. Thus, explaining the discrepancy through adiabatic
compression requires the baryonic fraction within the Einstein radius
to be $\sim 1/3$, twice the cosmic baryon fraction. This seems highly
unlikely, as the observed X-ray emission yields at these radii gas
fractions well below the cosmic value (e.g., see Figure~12 of
\citet{Doron} for A1689), and a cD galaxy contains only $\sim 1 \times
10^{12} M_\odot$ in baryons (e.g., see \citet{M87} for M87).

Modifications in the properties of dark matter or the slope of the
primordial power spectrum are generally expected to have a smaller
effect on clusters than on smaller-scale objects which are predicted
in $\Lambda$CDM to have earlier formation times and higher
concentrations. On the other hand, since clusters are rare objects in
the standard model, primordial non-Gaussianity would significantly
affect them and allow clusters to form earlier, which may also help
explain other observations \citep{silk,yoel}. Regardless of the
mechanism, early collapse of cluster cores may have observable
consequences if it is accompanied by star formation.

Finally, we note that the fact that clusters are now being detected
with masses $\sim 10^{15} M_\odot$ is completely consistent with the
$\Lambda$CDM model. Indeed, Figure~\ref{fig:Nofz} shows that large
numbers of clusters are expected out to significant redshifts,
including $M = 2 \times 10^{15} M_\odot$ halos out to $z \sim 1$, as
well as more massive halos up to $\sim 5\times 10^{15} M_\odot$ at
lower redshift. Clearly, while large samples of halos with precise,
profile-independent lensing determinations of both $\tE$ and $\Mv$
will make our results completely conclusive, the highly-significant
discrepancy we have identified already represents a substantial
challenge for $\Lambda$CDM.

\section*{Acknowledgments}
We thank Masataka Fukugita and Masahiro Takada for useful
discussions. RB is grateful for support from the ICRR in Tokyo, Japan
and from the John Simon Guggenheim Memorial Foundation. We acknowledge
Israel Science Foundation grants 629/05 (RB) and 1218/06 (TJB).


\label{lastpage}


\begin{thebibliography}{07}
 
\bibitem[\protect\citeauthoryear{Astier et al.}{2006}]{SNV06} Astier,
P., et al.\ 2006, A\&A, 447, 31

\bibitem[\protect\citeauthoryear{Bartelmann}{1996}]{Bart96}
Bartelmann, M.\ 1996, A\&A, 313, 697

\bibitem[\protect\citeauthoryear{Blumenthal et al.}{1986}]{blumenthal}
Blumenthal, G.~R., Faber, S.~M., Flores, R., \& Primack, J.~R.\ 1986,
ApJ, 301, 27

\bibitem[\protect\citeauthoryear{Brada{\v c} et
al.}{2007}]{Bradac} Brada{\v c}, M., et al.\ 2007, ArXiv
e-prints, 711, arXiv:0711.4850

\bibitem[\protect\citeauthoryear{Broadhurst et
al.}{2000}]{Br00} Broadhurst, T., Huang, X., Frye, B.,
\& Ellis, R.\ 2000, ApJ, 534, L15

\bibitem[\protect\citeauthoryear{Broadhurst et
al.}{2005a}]{Br05a} Broadhurst, T., et al.\ 2005a, ApJ,
621, 53

\bibitem[\protect\citeauthoryear{Broadhurst et
al.}{2005b}]{Br05b} Broadhurst, T., Takada, M., Umetsu,
K., Kong, X., Arimoto, N., Chiba, M., \& Futamase, T.\ 2005b, ApJ,
619, L143

\bibitem[\protect\citeauthoryear{Bryan \& Norman}{1998}]{bn98} Bryan
G.~L., \& Norman M., 1998, ApJ, 495, 80

\bibitem[\protect\citeauthoryear{Bullock et al.}{2000}]{Bull} Bullock
J.~S., Kolatt T.~S., Sigad Y., Somerville R.~S., Kravtsov A.~V.,
Klypin A.~A., Primack J.~R., \& Dekel A., 2000, MNRAS, 321, 559

\bibitem[\protect\citeauthoryear{Comerford et al.}{2006}]{Comerford}
Comerford, J.~M., Meneghetti, M., Bartelmann, M., \& Schirmer, M.\
2006, ApJ, 642, 39

\bibitem[\protect\citeauthoryear{Czoske et
al.}{2002}]{Czoske} Czoske, O., Moore, B., Kneib, J.-P.,
\& Soucail, G.\ 2002, A\&A, 386, 31

\bibitem[\protect\citeauthoryear{Diego et al.}{2005}]{Diego} Diego
J.~M., Sandvik H.~B., Protopapas P., Tegmark M., Ben{\'{\i}}tez N., \&
Broadhurst T.\ 2005, MNRAS, 362, 1247

\bibitem[\protect\citeauthoryear{Ford et al.}{1998}]{Ford} Ford,
H.~C., et al.\ 1998, procSPIE, 3356, 234

\bibitem[\protect\citeauthoryear{Gao et al.}{2007}]{Gao} Gao,
L., et al.\ 2007, MNRAS, submitted (arXiv:0711.0746)

\bibitem[\protect\citeauthoryear{Gavazzi et
al.}{2003}]{Gavazzi} Gavazzi, R., Fort, B., Mellier, Y.,
Pell{\'o}, R., \& Dantel-Fort, M.\ 2003, A\&A, 403, 11

\bibitem[\protect\citeauthoryear{Gioia et al.}{1990}]{Gioia} Gioia,
I.~M., Maccacaro, T., Schild, R.~E., Wolter, A., Stocke, J.~T.,
Morris, S.~L., \& Henry, J.~P.\ 1990, ApJS, 72, 567

\bibitem[\protect\citeauthoryear{Halkola et al.}{2008}]{halkola08}
Halkola A., Hildebrandt H., Schrabback T., Lombardi M., Bradac M.,
Erben T., Schneider P., Wuttke D., 2008, A\&A, in press

\bibitem[\protect\citeauthoryear{Halkola et
al.}{2006}]{Halkola} Halkola, A., Seitz, S., \& Pannella,
M.\ 2006, MNRAS, 372, 1425

\bibitem[\protect\citeauthoryear{Hennawi et al.}{2007}]{Hennawi} Hennawi 
J.~F., Dalal N., Bode P., \& Ostriker J.~P., 2007, ApJ, 654, 714

\bibitem[\protect\citeauthoryear{Jee et al.}{2007}]{Jee} Jee, M.~J.,
et al.\ 2007, ApJ, 661, 728

\bibitem[\protect\citeauthoryear{Kaiser et al.}{1995}]{KSB} Kaiser,
N., Squires, G., \& Broadhurst, T.\ 1995, ApJ, 449, 460

\bibitem[\protect\citeauthoryear{Kling et al.}{2005}]{kling} Kling,
T.~P., Dell'Antonio, I., Wittman, D., \& Tyson, J.~A.\ 2005, ApJ, 625,
643

\bibitem[\protect\citeauthoryear{Kneib et al.}{2003}]{Kneib} Kneib,
J.-P., et al.\ 2003, ApJ, 598, 804

\bibitem[\protect\citeauthoryear{Lemze et al.}{2007}]{Doron} Lemze D.,
Barkana R., Broadhurst T.~J., \& Rephaeli Y.\ 2007, MNRAS, accepted

\bibitem[\protect\citeauthoryear{Limousin et al.}{2007}]{Limousin}
Limousin, M., et al.\ 2007, ApJ, 668, 643

\bibitem[\protect\citeauthoryear{Mathis et al.}{2004}]{silk} Mathis
H., Diego J.~M., \& Silk J.\ 2004, MNRAS, 353, 681

\bibitem[\protect\citeauthoryear{Matsushita et al.}{2002}]{M87}
Matsushita, K., Belsole, E., Finoguenov, A., B\"{o}hringer, H.\
2002, A\&A, 386, 77

\bibitem[\protect\citeauthoryear{Medezinski et
al.}{2007}]{Medezinski} Medezinski, E., et al.\ 2007, ApJ,
663, 717

\bibitem[\protect\citeauthoryear{Navarro et al.}{1997}]{NFW} Navarro
J.~F., Frenk C.~S., \& White S.~D.~M.\ 1997, ApJ, 490, 493 (NFW)

\bibitem[\protect\citeauthoryear{Neto et al.}{2007}]{Neto} Neto,
A.~F., et al.\ 2007, MNRAS, 381, 1450

\bibitem[\protect\citeauthoryear{Oguri et al.}{2005}]{Oguri} Oguri
M., Takada M., Umetsu K., \& Broadhurst T.\ 2005, ApJ, 632, 841

\bibitem[\protect\citeauthoryear{Percival et al.}{2007}]{BAO07}
Percival, W.~J., Cole, S., Eisenstein, D.~J., Nichol, R.~C., Peacock,
J.~A., Pope, A.~C., \& Szalay, A.~S.\ 2007, MNRAS, 381, 1053

\bibitem[\protect\citeauthoryear{Sadeh et al.}{2007}]{yoel} Sadeh S.,
Rephaeli Y., \& Silk J.\ 2007, MNRAS, 380, 637

\bibitem[\protect\citeauthoryear{Sand et al.}{2005}]{Sand} Sand
D.~J., Treu T., Ellis R.~S., \& Smith G.~P.\ 2005, ApJ, 627, 32

\bibitem[\protect\citeauthoryear{Sheth, Mo, \& Tormen}{2001}]{Sheth} 
Sheth R.~K., Mo H.~J., \& Tormen G.\ 2001, MNRAS, 323, 1

\bibitem[\protect\citeauthoryear{Smith et al.}{2005}]{Smith} Smith,
G.~P., Kneib, J.-P., Smail, I., Mazzotta, P., Ebeling, H., \& Czoske,
O.\ 2005, MNRAS, 359, 417

\bibitem[\protect\citeauthoryear{Spergel et al.}{2007}]{Spergel07}
  Spergel D.~N. et al.\ 2007, ApJS, 170, 377

\bibitem[\protect\citeauthoryear{Umetsu \& Broadhurst}{2008}]{umetsu}
Umetsu, K., \& Broadhurst, T.\ 2008, submitted (arXiv:0712.3441)

\bibitem[\protect\citeauthoryear{Vikhlinin et al.}{2002}]{vik}
Vikhlinin, A., VanSpeybroeck, L., Markevitch, M., Forman, W.~R., \&
Grego, L.\ 2002, ApJ, 578, L107

\bibitem[\protect\citeauthoryear{Williams et al.}{1999}]{navarro}
Williams, L.~L.~R., Navarro, J.~F., \& Bartelmann, M.\ 1999, ApJ,
527, 535

\bibitem[\protect\citeauthoryear{Zekser et al.}{2006}]{Zekser} 
Zekser K.~C. et al.\ 2006, ApJ, 640, 639 

\bibitem[\protect\citeauthoryear{Zhao et al.}{2003}]{Jing} Zhao,
D.~H., Jing, Y.~P., Mo, H.~J., B\"{o}rner, G.\ 2003, ApJ, 597,
L9

\end{thebibliography}
\end{document}